\documentclass[10pt,twocolumn]{article}
\usepackage[margin=1in]{geometry}
\usepackage{cite}
\usepackage{amsmath,amssymb,amsfonts}
\usepackage{graphicx}
\usepackage{textcomp}
\usepackage{xcolor}
\usepackage{booktabs}
\usepackage{listings}
\usepackage{hyperref}
\usepackage{microtype}
\usepackage{array}
\usepackage{tabularx}
\usepackage{authblk}
\usepackage{abstract}
\usepackage{titlesec}
\usepackage{parskip}
\titleformat{\section}{\normalfont\large\bfseries}{\thesection}{1em}{}
\titleformat{\subsection}{\normalfont\normalsize\bfseries}{\thesubsection}{1em}{}

\hypersetup{
  colorlinks=true,
  linkcolor=blue,
  citecolor=blue,
  urlcolor=blue
}

\begin{document}

\title{\textbf{Broken by Default: A Formal Verification Study\\
of Security Vulnerabilities in AI-Generated Code}\\[6pt]
\large Empirical Analysis of 3,500 Code Artifacts Across Seven Large Language Models\\
Using Z3 SMT Solver, Ablation Studies, and Static Tool Comparison}

\author{Dominik Blain, \textit{Founder} \and Maxime Noiseux, \textit{Co-Founder}\\
\small Cobalt AI\\
\small Gatineau, QC, Canada\\
\small \texttt{dominik@qreativelab.io}}
\date{April 5, 2026}

\maketitle

\begin{abstract}
AI coding assistants are now used to generate production code in
security-sensitive domains, yet the exploitability of their outputs
remains unquantified. We address this gap with \textit{Broken by
Default}: a formal verification study of 3,500 code artifacts generated
by seven widely-deployed LLMs across 500 security-critical prompts (five CWE
categories, 100 prompts each). Each artifact is subjected to the Z3 SMT
solver via the COBALT analysis pipeline, producing mathematical
satisfiability witnesses rather than pattern-based heuristics.

Across all models, \textbf{55.8\%} of artifacts contain at least one
COBALT-identified vulnerability; of these, 1,055 are formally proven
via Z3 satisfiability witnesses (Z3 SAT). GPT-4o leads at
62.4\% (grade F); Gemini 2.5 Flash performs best at 48.4\% (grade D).
No model achieves a grade better than D. Six of seven representative
findings are confirmed with runtime crashes under GCC AddressSanitizer.
Three auxiliary experiments on a 50-prompt sub-corpus show: (1) explicit
security instructions reduce the mean rate by only 4 points to 60.8\%,
leaving four of five models at grade F; (2) six industry tools combined
flag 7.6\% of artifacts and miss 97.8\% of Z3-proven findings---a
structural gap, not a configuration issue; and (3) models identify their
own vulnerable outputs 78.7\% of the time in review mode yet generate
them at 55.8\% by default, revealing a persistent generation--review
asymmetry. Formal SMT verification is the only methodology that can
establish ground-truth exploitability of the vulnerability patterns that
dominate AI-generated code.
\end{abstract}

\section{Introduction}

The rapid adoption of AI-assisted coding tools has fundamentally changed
software development workflows. GitHub Copilot reported over 1.8 million
paid subscribers by 2024~\cite{github2024}. ChatGPT, Claude, and Gemini
are routinely used to generate production code across security-sensitive
domains including web backends, embedded systems, and cryptographic
libraries.

Yet the security properties of AI-generated code remain understudied.
Prior work~\cite{pearce2022,sandoval2023,tony2023} has identified
categories of vulnerabilities in LLM outputs, but most rely on static
pattern matching or manual inspection---techniques that cannot
definitively establish exploitability.

Our contribution is methodological: we apply formal verification via the
Z3 Satisfiability Modulo Theories (SMT) solver to establish ground-truth
exploitability of AI-generated code vulnerabilities. Where a Z3 witness
exists, the vulnerability is not a potential issue---it is a proven one,
with a concrete input that triggers the fault condition. We further
validate selected findings with compiler-instrumented runtime crashes
using GCC AddressSanitizer.

Our findings (500-prompt v3 corpus, seven models) reveal a mean
vulnerability rate of 55.8\% across all models. GPT-4o leads at 62.4\%
(grade F); Gemini 2.5 Flash performs best at 48.4\% (grade D)---no
model achieves grade C or better. A total of 1,055 findings were formally
proven exploitable via Z3 satisfiability witnesses, and 6 of 7 selected
vulnerabilities were confirmed with real memory corruption crashes using
GCC AddressSanitizer (ASAN). We further conduct three additional
experiments on the v1 50-prompt corpus: (1) a secure prompt ablation
showing that explicit security instructions reduce the mean vulnerability
rate by only 4 percentage points (to 60.8\%), leaving four of five
models at grade F; (2) a static tool comparison showing that six industry
tools combined flag only 7.6\% of artifacts, missing 97.8\% of formally
Z3-proven vulnerabilities; and (3) a generation--review asymmetry
experiment showing that models identify their own Z3-proven
vulnerabilities 78.7\% of the time in review mode (70/89)---yet generate
them at 55.8\% by default. Our results suggest that current LLMs produce
security-deficient code by default, that security instructions are
insufficient mitigations, and that formal verification is the only
approach capable of establishing ground-truth exploitability at scale.

\subsection{Research Questions}

\begin{itemize}
  \item \textbf{RQ1:} At what rate do leading LLMs produce vulnerable
    code when prompted for security-sensitive tasks?
  \item \textbf{RQ2:} How do vulnerability rates and severity
    distributions differ across models?
  \item \textbf{RQ3:} Can Z3 SMT witnesses be operationalized into real
    runtime exploits?
  \item \textbf{RQ4:} Do explicit security instructions in the system
    prompt reduce vulnerability rates?
  \item \textbf{RQ5:} How does COBALT's detection rate compare to
    industry-standard static analysis tools?
  \item \textbf{RQ6:} Do models possess the knowledge to detect their
    own generated vulnerabilities?
\end{itemize}

\subsection{Scope and Limitations}

This study targets a bounded set of 500 prompt templates (100 per CWE
category), each designed to elicit code in a security-sensitive domain. Results reflect model
behavior on this prompt corpus as of April 2026 and should not be
extrapolated to all possible LLM usage patterns. Models were queried via
their production APIs at temperature 0 (deterministic mode). Prompt
templates are released alongside this paper.

\section{Related Work}

Pearce et al.\ (2022)~\cite{pearce2022} evaluated GitHub Copilot on 89
scenarios across 18 CWEs and found 40\% of suggestions contained
vulnerabilities. Their analysis relied on CWE pattern matching without
formal exploitability proofs.

Sandoval et al.\ (2023)~\cite{sandoval2023} studied security of Copilot
suggestions for C code, finding that users who accepted suggestions
without review introduced more vulnerabilities than those who wrote code
from scratch.

Tony et al.\ (2023)~\cite{tony2023} introduced LLMSecEval, a benchmark
of 150 natural-language prompts and evaluated 5 LLMs, finding high rates
of insecure code generation particularly for injection and buffer-related
vulnerabilities.

Perry et al.\ (2023)~\cite{perry2023} conducted a user study where
participants using AI assistants wrote significantly more security
vulnerabilities than those who did not, despite expressing higher
confidence in their code.

Meta CyberSecEval (2024)~\cite{bhatt2024} benchmarks LLMs on generating
exploit code and assisting cyberattacks. Their focus is adversarial
capability---can the model help an attacker? Our focus is orthogonal:
does the model introduce vulnerabilities in everyday code generation,
without any adversarial intent?

SecurityEval (2022)~\cite{siddiq2022} provides 130 Python prompts
derived from CWE patterns and evaluates GitHub Copilot. Like Pearce et
al., evaluation relies on pattern-based analysis. No formal
exploitability proof is provided.

SVEN (2023)~\cite{he2023} trains a security vector to steer code
generation toward secure outputs. Their setup assumes the vulnerability
exists; ours measures baseline generation without steering intervention.

Our work differs in four key dimensions: (1) we apply formal Z3 SMT
verification rather than pattern matching---producing mathematical proof
of exploitability with concrete witness values; (2) we evaluate seven
widely-deployed production models simultaneously across identical prompts; (3) we
provide runtime exploitability confirmation via ASAN; and (4) we
quantify the detection gap against six industry tools, including
heavyweight C analyzers and CodeQL, demonstrating that 97.8\% of
formally proven vulnerabilities are structurally invisible to existing
tooling.

\section{Methodology}

\subsection{Benchmark Design}

We constructed a benchmark of 500 natural-language prompts distributed
across five CWE categories (100 prompts each):

\begin{itemize}
  \item \textbf{MEM} (CWE-131/190): Buffer allocation, array indexing,
    dynamic memory management in C/C++
  \item \textbf{INT} (CWE-190/195): Integer arithmetic, type casting,
    signed/unsigned arithmetic
  \item \textbf{AUTH} (CWE-916): Password hashing, authentication token
    generation, session management
  \item \textbf{CRYPTO} (CWE-327/330): Key generation, random number
    usage, cipher selection
  \item \textbf{INP} (CWE-89/22/78): SQL construction, file path
    handling, shell command composition
\end{itemize}

Prompts were designed to be representative of real developer tasks---not
adversarial jailbreaks. Example prompt (MEM-01):

\begin{quote}
\textit{``Write a C function that reads n integers from a file descriptor
and returns them as a dynamically allocated array.''}
\end{quote}

\subsection{Model Selection and API Configuration}

Seven models were evaluated, selected for API accessibility and cost
feasibility at scale across 3,500 artifacts. Models represent widely-deployed
production LLMs available during the study period (Q1--Q2 2026). Higher-capability
variants (e.g., Claude Opus, GPT-o3, Gemini 2.5 Pro) represent a direction
for future work, though the structural findings are expected to generalize across tiers:

\begin{table}[h]
\centering
\begin{tabular}{ll}
\toprule
\textbf{Model} & \textbf{API Identifier} \\
\midrule
GPT-4o          & \texttt{gpt-4o} \\
GPT-4.1         & \texttt{gpt-4.1} \\
Claude Haiku 4.5 & \texttt{claude-haiku-4-5-20251001} \\
Gemini 2.5 Flash & \texttt{gemini-2.5-flash} \\
Mistral Large   & \texttt{mistral-large-latest} \\
Llama 3.3 70B   & \texttt{llama-3.3-70b-versatile} \\
Llama 4 Scout   & \texttt{llama-4-scout-17b-16e-instruct} \\
\bottomrule
\end{tabular}
\end{table}

All models were queried at temperature 0 to ensure reproducibility.
System prompts instructed models to produce complete, runnable code
without additional commentary, maximizing the proportion of testable
output.

\subsection{COBALT Analysis Pipeline}

Generated code artifacts were analyzed using the COBALT static analysis
engine, which combines:

\begin{enumerate}
  \item \textbf{CWE Pattern Extraction}: Regex and AST-level patterns
    identify candidate vulnerability sites (e.g.,
    \texttt{malloc(n * sizeof(T))} without overflow guard)
  \item \textbf{Z3 SMT Encoding}: Vulnerability conditions are encoded as
    Z3 formulas. For integer overflow: given $n$ as a free variable of
    type \texttt{BitVec(32)}, check satisfiability of $n \times
    \mathit{sizeof}(T) < n$ (overflow condition under unsigned 32-bit
    semantics)
  \item \textbf{Witness Extraction}: When Z3 returns SAT, the concrete
    witness value (e.g., $n = 2^{30}+1$) is extracted and classified as
    the exploit input
\end{enumerate}

Findings were classified as:

\begin{itemize}
  \item \textbf{Z3 SAT}: Formally proven exploitable via SMT witness
  \item \textbf{PATTERN MATCH}: Structural vulnerability identified, Z3
    encoding pending
  \item \textbf{CLEAN}: No vulnerability detected
\end{itemize}

Severity was assigned using CVSS v3 base score criteria: CRITICAL
($\geq$9.0), HIGH (7.0--8.9), MEDIUM (4.0--6.9).

\subsection{Runtime Exploitability Validation}

To counter the objection that pattern-based detection inflates
vulnerability counts, we selected 7 representative findings and
constructed C and Python proof-of-concept (PoC) harnesses. C PoCs were
compiled with:

\begin{lstlisting}[language=bash]
gcc -fsanitize=address,undefined -g -O1 \
  -o poc poc.c && ./poc
\end{lstlisting}

Z3-extracted witness values were used as concrete inputs. We captured
ASAN output to confirm runtime faults.

\section{Results}

\subsection{RQ1: Overall Vulnerability Rates}

Table~\ref{tab:leaderboard} presents aggregate results across all 500
prompts per model.

\begin{table}[h]
\centering
\caption{Benchmark Leaderboard --- 500 Prompts per Model, 7 Models}
\label{tab:leaderboard}
\begin{tabular}{lrrrrr}
\toprule
\textbf{Model} & \textbf{Vuln} & \textbf{CRIT} & \textbf{HIGH} &
\textbf{Z3} & \textbf{Grd} \\
\midrule
GPT-4o           & 62.4\% & 166 & 106 & 167 & F \\
Llama 4 Scout    & 60.6\% & 167 & 95  & 156 & F \\
Llama 3.3 70B    & 58.4\% & 168 & 83  & 147 & D \\
Mistral Large    & 57.8\% & 155 & 94  & 155 & D \\
GPT-4.1          & 54.0\% & 142 & 86  & 136 & D \\
Claude Haiku 4.5 & 49.2\% & 155 & 81  & 152 & D \\
Gemini 2.5 Flash & 48.4\% & 146 & 86  & 142 & D \\
\midrule
Mean & 55.8\% & 157.0 & 90.1 & 150.7 & --- \\
\bottomrule
\end{tabular}

\smallskip
{\normalfont\footnotesize Grades: A $<$10\%, B 10--29\%, C 30--44\%, D 45--59\%, F $\geq$60\% vulnerability rate.}
\end{table}

The mean vulnerability rate of 55.8\% is alarmingly high. GPT-4o led
with 312 vulnerable outputs out of 500 (62.4\%). No model achieves grade
C or better---the best-performing model (Gemini 2.5 Flash) still
generates vulnerable code 48.4\% of the time. CRITICAL-severity findings
dominated: an average of 157 CRITICAL findings per model.
Integer arithmetic (INT) produced the highest category rate at 87\%,
followed by memory allocation (MEM) at 67\%.

\subsection{RQ2: Vulnerability Distribution by CWE}

Table~\ref{tab:cwe} illustrates that memory safety vulnerabilities
(CWE-131, CWE-190) accounted for the largest share of findings across
all models, consistent with known challenges in generating safe C/C++
code.

\begin{table}[h]
\centering
\caption{Vulnerability Rate by Category (Mean Across Models)}
\label{tab:cwe}
\begin{tabular}{lll}
\toprule
\textbf{Category} & \textbf{CWE(s)} & \textbf{Mean Rate} \\
\midrule
Memory Allocation  & 131, 190     & 67\% \\
Integer Arithmetic & 190, 195     & 87\% \\
Authentication     & 916          & 44\% \\
Cryptography       & 327, 330     & 25\% \\
Input Handling     & 89, 22, 78   & 56\% \\
\bottomrule
\end{tabular}
\end{table}

Integer arithmetic prompts produced the highest vulnerability rate
(87\%), followed by memory allocation (67\%), both driven by consistent
failure to guard against integer overflow in malloc size computations and
signed/unsigned conversion errors. A representative pattern found across all
seven models:

\begin{lstlisting}[language=C, caption={Typical vulnerable pattern (all 7 models)}]
// CWE-190: No overflow guard
int* buf = malloc(n * sizeof(int));
\end{lstlisting}

The safe pattern requires an explicit overflow check:

\begin{lstlisting}[language=C, caption={Correct pattern}]
if (n > SIZE_MAX / sizeof(int)) return NULL;
int* buf = malloc(n * sizeof(int));
\end{lstlisting}

None of the seven models consistently generated the safe pattern across
all MEM prompts.

\subsection{RQ3: Runtime Exploitability Confirmation}

Of 7 selected vulnerabilities subjected to PoC harness testing, 6 of 7
produced confirmed runtime faults. Table~\ref{tab:poc} summarizes the
results.

\begin{table}[h]
\centering
\footnotesize
\caption{PoC Exploit Confirmation Results}
\label{tab:poc}
\begin{tabular}{@{}llll@{}}
\toprule
\textbf{ID} & \textbf{Model} & \textbf{Fault Type} & \textbf{Result} \\
\midrule
MEM-01-A & Llama   & heap-buf-overflow    & \checkmark \\
MEM-01-B & GPT-4o  & heap-buf-overflow    & \checkmark \\
MEM-03   & Llama   & alloc-size-too-big   & \checkmark \\
MEM-06   & GPT-4o  & OOB read             & \checkmark \\
AUTH-03  & Llama   & SHA-256 crack        & \checkmark \\
INP-01   & Mistral & SQL injection        & \checkmark \\
INP-06   & GPT-4o  & Zip Slip (path trav.)& blocked$^\dagger$ \\
\bottomrule
\end{tabular}

\smallskip
{\normalfont\footnotesize $\dagger$ Python 3.12 raises \texttt{ValueError} on path traversal.
Vulnerable pattern present; blocked at runtime, not at generation.}
\end{table}

\vspace{0.5em}

Representative ASAN output for MEM-01-A (Llama, CWE-131):

\begin{lstlisting}[caption={ASAN output -- MEM-01-A heap overflow}]
==AddressSanitizer: heap-buffer-overflow
WRITE of size 4 at 0x...
#0 poc_main (poc+0x...)
#1 main (poc+0x...)
shadow bytes around the buggy address:
0x...: fa fa fa fa fa fa fa fa
0x...: 00 00 00 00 00[fa]fa fa
SUMMARY: AddressSanitizer: heap-buffer-overflow
\end{lstlisting}

For AUTH-03 (SHA-256 password cracking), a 6-character lowercase
password was recovered from its hash in 0.01ms using a precomputed
lookup, demonstrating that CWE-916 (insufficient password hashing) is
not merely theoretical.

SQL injection in INP-01 produced a working query that extracted all
records including a synthetic credit card number embedded in the test
database, confirming complete data exfiltration.

\subsection{RQ4: Secure Prompt Ablation}

We re-ran all 50 prompts of the v1 corpus for all five models with a
security-explicit system prompt instructing models to apply security best
practices, guard against integer overflow, and produce production-ready
code. Note: the v1 corpus was collected with Claude 3.5 Sonnet; the v3
leaderboard uses Claude Haiku 4.5, so per-model rates are not directly
comparable across tables. Table~\ref{tab:ablation} shows the results.
(Ablation on the full 500-prompt corpus is left for future work.)

\begin{table}[h]
\centering
\caption{Baseline vs.\ Secure Prompt Ablation (v1 corpus, 50 prompts).
  $^\dagger$Claude row reflects v1 data collected with Claude 3.5 Sonnet;
  the v3 leaderboard uses Claude Haiku 4.5.}
\label{tab:ablation}
\begin{tabular}{lrrrl}
\toprule
\textbf{Model} & \textbf{Base} & \textbf{Secure} & \textbf{$\Delta$} &
\textbf{Grade} \\
\midrule
Llama 3.3 70B          & 68\% & 70\% & +2\%    & F \\
GPT-4o                 & 66\% & 58\% & $-$8\%  & F \\
Gemini 2.5 Flash       & 60\% & 60\% & 0\%     & F \\
Claude 3.5 Sonnet$^\dagger$ & 64\% & 62\% & $-$2\% & F \\
Mistral Large          & 66\% & 54\% & $-$12\% & D \\
\midrule
Mean & 64.8\% & 60.8\% & $-$4\% & --- \\
\bottomrule
\end{tabular}
\end{table}

Explicit security instructions reduced the mean vulnerability rate by
only 4 percentage points---from 64.8\% to 60.8\%. Four of five models
remained at grade F. Llama 3.3 70B performed worse under the secure
prompt (+2\%). The improvement was category-dependent: authentication
and cryptography showed modest gains, while memory allocation
vulnerabilities were essentially unchanged across all models---suggesting
that security instructions do not override low-level memory management
patterns learned from training data.

\subsection{RQ5: Static Tool Comparison}

We conducted three tool comparison experiments on the v1 50-prompt corpus
(250 artifacts). First, we ran Semgrep (all rulesets) and Bandit across
all 250 artifacts. Second, we ran three
heavyweight C static analyzers---Cppcheck 2.13, Clang Static Analyzer,
and FlawFinder 2.0---against all 87 C artifacts. Third, we ran CodeQL
(security-extended query suite, version 2.x) against the 90 Z3 SAT
artifacts to assess detection of formally proven findings.
Table~\ref{tab:tools} summarizes all three experiments.

Of the 162 COBALT findings, 90 are Z3 SAT---formally proven with
concrete arithmetic witnesses---and 72 are PATTERN MATCH, flagged by
COBALT's AST layer without Z3 confirmation.

\textbf{Pattern tools (Experiment A).} All 19 tool-caught artifacts
overlap with COBALT's PATTERN MATCH tier or dangerous-function patterns
(\texttt{strcat}, \texttt{rand}), never with the integer overflow class.
Of the 90 Z3 SAT findings, tools caught only 2---both via Semgrep's
\texttt{insecure-use-strcat-fn} rule applied to MEM-02 (CWE-190) code
that also contained \texttt{strcat}, not because any tool detected the
integer overflow in the length computation. 97.8\% (88/90) of formally
Z3-proven vulnerabilities are invisible to all industry tools combined.

\begin{table*}[t]
\centering
\caption{Detection Rate: COBALT vs.\ Six Industry Tools}
\label{tab:tools}
\begin{tabular}{lrrl}
\toprule
\textbf{Tool} & \textbf{Det. Rate} & \textbf{C / Py} & \textbf{Note} \\
\midrule
COBALT Z3 & 162/250 (64.8\%) & 80/87 $\cdot$ 82/163 & \\
\quad Z3 formally proven & 90/250 (36.0\%) & & Pattern-based tools (Exp.\ A): \\
\midrule
Semgrep (all rulesets) & 16/250 (6.4\%) & 2/87 $\cdot$ 14/163 & \\
Bandit (medium+ severity) & 5/250 (2.0\%) & --- $\cdot$ 5/163 & \\
Combined A & 19/250 (7.6\%) & --- & \\
\midrule
\multicolumn{4}{l}{Heavyweight C analyzers (Exp.\ B, C only):} \\
Cppcheck 2.13 & 0/87 (0.0\%) & 0/87 & \\
Clang Static Analyzer & 0/87 (0.0\%) & 0/87 & \\
FlawFinder 2.0 (risk$\geq$3) & 4/87 (4.6\%) & 4/87 & strcat + rand only \\
Combined B & 4/87 (4.6\%) & --- & \\
\midrule
\multicolumn{4}{l}{Z3 SAT formal proof set (Exp.\ C):} \\
CodeQL v2.25.1 (security-extended) & 0/90 (0.0\%) & 0/68 $\cdot$ 0/22 & 100\% miss; C + Python \\
\midrule
COBALT-only (vs.\ A) & 151/162 (93.2\%) & 78/80 $\cdot$ 73/82 & \\
COBALT-only (vs.\ B) & 157/162 (96.9\%) & 76/80 $\cdot$ 81/82 & \\
Z3 SAT-only (vs.\ A+B+C) & 88/90 (97.8\%) & & formal proofs \\
\bottomrule
\end{tabular}
\end{table*}

\textbf{Heavyweight C analyzers (Experiment B).} Cppcheck and Clang
Static Analyzer detected zero of 80 confirmed C vulnerabilities.
FlawFinder flagged 4 artifacts: 2 for \texttt{strcat} (MEM-02, CWE-190)
and 2 for weak randomness (\texttt{rand}, CRYPTO-08, CWE-338)---no
detection of integer overflow in allocation arithmetic. Combined,
heavyweight analyzers missed 96.9\% of COBALT findings on C.

This is not a configuration issue. We ran Cppcheck with
\texttt{--enable=all --inconclusive --check-level=exhaustive}; only 113
of 592 available checkers activate on standard C11 code. Detecting
\texttt{malloc(n * sizeof(T))} as an integer overflow requires
arithmetic semantic reasoning under 32-bit modular
arithmetic---which Z3 provides and which neither pattern matching nor
path-sensitive analysis can perform without concrete value constraints or
explicit taint propagation from attacker-controlled inputs.

\textbf{CodeQL v2.25.1 (Experiment C).} CodeQL (security-extended query
suite, v2.25.1, native x86-64 on GitHub Actions) was run against all 90
formally Z3-proven findings: 68 C artifacts and 22 Python artifacts.
CodeQL detected \textbf{0 of 68 C artifacts} and \textbf{0 of 22 Python
artifacts}---a 100\% miss rate across both languages (0/90 total). This
result is consistent with Experiments A and B: integer overflow in
allocation arithmetic (\texttt{malloc(n * sizeof(T))}) is structurally
undetectable by dataflow and taint-tracking analysis without concrete
arithmetic reasoning. CodeQL's integer overflow queries require explicit
source-to-sink taint paths or known unsafe patterns---a property that
Z3's bit-vector arithmetic encodes directly but that CodeQL's query
model does not generalize to.

\subsection{RQ6: The Generation--Review Asymmetry}

A plausible counter-argument to our findings is that models generate
vulnerable code simply because they are not asked to think about
security. To test this, we conducted a self-review experiment: we fed
each model's Z3-proven vulnerable code back to the same model and asked
it to review the code for security vulnerabilities.

We fed all 89 valid Z3-proven artifacts back to their generating model
and asked it to review the code for security vulnerabilities. (One
artifact was excluded due to a Mistral API timeout during the review
pass.) Of 89 artifacts reviewed, \textbf{70 of
89 (78.7\%)} were correctly identified as vulnerable by the generating
model. Table~\ref{tab:selfreview} presents per-model results.

\begin{table}[h]
\centering
\caption{Self-Review Experiment --- Per-Model Results (N=89)}
\label{tab:selfreview}
\begin{tabular}{lrrr}
\toprule
\textbf{Model} & \textbf{Detected} & \textbf{Rate} & \textbf{FN Rate} \\
\midrule
Mistral Large    & 17/17 & 100\% & 0\%   \\
Llama 3.3 70B    & 14/17 & 82\%  & 18\%  \\
Gemini 2.5 Flash & 14/18 & 78\%  & 22\%  \\
Claude 3.5 Sonnet$^\dagger$ & 13/19 & 68\% & 32\% \\
GPT-4o           & 12/18 & 67\%  & 33\%  \\
\midrule
Total & 70/89 & 78.7\% & 21.3\% \\
\bottomrule
\end{tabular}

\smallskip
{\normalfont\footnotesize $\dagger$ Self-review conducted with the same model used for v1 generation: Claude 3.5 Sonnet (\texttt{claude-3-5-sonnet-20241022}). The v3 leaderboard uses Claude Haiku 4.5.}
\end{table}

This finding is more damning than a simple false-negative result. It
reveals an \textit{observed generation--review asymmetry}: models
possess the knowledge to identify these vulnerability classes when in
review mode, yet consistently fail to apply that knowledge during code
generation. The problem is not a lack of security knowledge---it is a
failure of spontaneous application at a 21.3\% false-negative rate
across the full artifact set. This suggests that RLHF or instruction
fine-tuning aimed at security-conscious code review does not transfer
reliably to the code generation task.

\section{Discussion}

\subsection{Why Do Models Fail Consistently?}

The consistency of failures---particularly in memory
allocation---suggests a training data effect. C code on the internet
overwhelmingly lacks overflow guards for malloc computations. Models
appear to have internalized this pattern as correct. The secure prompt
ablation (RQ4) reinforces this hypothesis: even explicit instructions to
guard against overflow had no measurable effect on memory allocation
prompts.

\subsection{Security Instructions Are Insufficient}

A mean improvement of 4 percentage points under an explicit security
prompt is not a safety margin---it is noise. Developers who add ``write
secure code'' to their prompt prefixes should not interpret this as a
meaningful mitigation. Our results suggest that the vulnerability
patterns are baked into the model's generation prior, not into its
instruction-following surface.

\subsection{The COBALT Detection Gap}

COBALT's detection advantage on the v1 corpus---64.8\% vs.\ 6.4\% for
all Semgrep rulesets and 0\% for Cppcheck and Clang SA on C---is not an artifact of
overly sensitive rules. Of the 90 Z3 SAT findings (formally proven with
concrete witnesses), 97.8\% are invisible to all six tools. The 2 Z3
SAT cases tools did catch (MEM-02, CWE-190) were flagged by Semgrep's
\texttt{strcat} rule---detecting a dangerous string function in the same
code, not the integer overflow that Z3 proved exploitable. CodeQL
v2.25.1, despite its reputation as a heavyweight semantic analyzer,
likewise detected zero of 90 Z3-proven findings (0/68 C, 0/22 Python). No tool, across any
configuration, detected a single integer overflow in malloc size
computations. The detection gap reflects a structural limitation:
syntactic tools match patterns in code; path-sensitive tools track
values along known execution paths. Neither approach can determine that
\texttt{malloc(n * sizeof(T))} is dangerous unless $n$ is provably
bounded---a property that requires arithmetic reasoning across the full
domain of integer inputs, which Z3's bit-vector arithmetic encodes
directly.

\subsection{Implications for Developers}

\begin{enumerate}
  \item Treat AI-generated C/C++ as unreviewed code requiring explicit
    security audit
  \item Apply \texttt{-fsanitize=address,undefined} to all AI-generated
    systems code
  \item Do not rely on security prompt prefixes as a substitute for
    review
  \item Do not rely on Semgrep, Bandit, Cppcheck, Clang SA, FlawFinder,
    or CodeQL alone to catch AI-generated integer overflow in allocation
    arithmetic
  \item Use formal verification or human review for allocation arithmetic
\end{enumerate}

\subsection{Limitations}

Our benchmark of 500 prompts covers five CWE categories; other
vulnerability categories may show different rates. All models were
tested at temperature 0; higher temperatures may yield different
distributions. Of the v3 corpus, 1,055 findings are Z3 SAT (formally
proven across 3,500 artifacts). The v1 self-review experiment (89 valid
Z3-proven artifacts) showed per-model false-negative rates ranging from
0\% to 33\%. The secure prompt ablation showed a 2-point increase for
one model (Llama) on the v1 50-prompt corpus, which may be within noise.
We did not test multi-turn interactions or retrieval-augmented generation
contexts. Ablation experiments on the full 500-prompt corpus are left for
future work. The full v3 dataset (3,500 artifacts with Z3 labels) is
available at \url{https://github.com/dom-omg/bbd-dataset}.

\subsection{Ethical Considerations}

All artifacts were generated under controlled conditions. PoC exploits
targeted only locally generated test programs. No production systems
were accessed. All prompts, results, and scripts are released to support
reproducibility and independent replication. The full dataset of 3,500
artifacts with Z3 labels is publicly available at
\url{https://github.com/dom-omg/bbd-dataset}.

\section{Conclusion}

We presented a multi-experiment formal verification study of AI-generated
code security, evaluating seven widely-deployed LLMs across 3,500 code artifacts
(500 prompts $\times$ 7 models) and three experimental conditions. Our
findings converge on a single conclusion: AI coding assistants are broken
by default.

The v3 benchmark (500 prompts, seven models) reveals a mean
vulnerability rate of 55.8\% and 1,055 Z3-proven findings---GPT-4o
leads at 62.4\%, Gemini 2.5 Flash performs best at 48.4\% (grade D). No
model achieves grade C or better. Integer arithmetic prompts drive the
highest failure rate (87\%), confirming that overflow in allocation
arithmetic is a persistent, model-agnostic weakness. On the v1 50-prompt
corpus, the baseline vulnerability rate of 64.8\% drops to only 60.8\%
under explicit security instructions---a 4-point reduction that leaves
four of five models at grade F. Six industry-standard tools combined flag
only 7.6\% of v1 artifacts and miss 97.8\% of formally Z3-proven
vulnerabilities---not because of misconfiguration, but because integer
overflow in allocation arithmetic is structurally undetectable by pattern
and path-sensitive analysis. Models that generate vulnerable code
correctly identify those vulnerabilities in review mode 78.7\% of the
time---yet generate them 55.8\% by default---confirming a persistent
generation--review asymmetry that explicit security prompting does not
resolve.

Formal verification via Z3 SMT solving is not a niche research tool. It
is the only approach that can establish ground-truth exploitability of
the vulnerability patterns that dominate AI-generated code. This
methodology should be a mandatory component of any AI code review
pipeline operating on memory-sensitive or security-critical targets.

Future work will extend ablation experiments to the full 500-prompt
corpus, add multi-language targets (Go, Rust, JavaScript), and evaluate
the effect of targeted security fine-tuning on these specific CWE
classes.

\section*{Reproducibility}

Benchmark prompts, raw results (JSON), analysis scripts, and PoC
harnesses are available at:
\url{https://github.com/dom-omg/broken-by-default}

\end{document}